\documentclass[9pt,conference]{IEEEtran}
\IEEEoverridecommandlockouts

\usepackage{cite}
\usepackage{amsmath,amssymb,amsfonts}
\usepackage{algorithmic}
\usepackage{graphicx}
\usepackage{textcomp}
\usepackage{multirow}
\usepackage{booktabs}
\usepackage{stfloats}
\usepackage{makecell}
\usepackage{hyperref}
\usepackage{float}
\usepackage{bbm}
\usepackage[dvipsnames]{xcolor}

\def\BibTeX{{\rm B\kern-.05em{\sc i\kern-.025em b}\kern-.08em
    T\kern-.1667em\lower.7ex\hbox{E}\kern-.125emX}}

\newcommand{\hl}[1]{\underline{\textbf{#1}}}

\begin{document}

\title{
    Disentangling Speakers in Multi-Talker Speech Recognition with Speaker-Aware CTC
}

\author{

    \Large{\textit{Jiawen Kang, Lingwei Meng, Mingyu Cui, Yuejiao Wang, Xixin Wu, Xunying Liu, Helen Meng}}
    
    \vspace{4pt}\\
    
    \Large{The Chinese University of Hong Kong, Hong Kong SAR, China} 

    % \vspace{-1pt}
}

% \author{\IEEEauthorblockN{1\textsuperscript{st} Jiawen Kang}
% \IEEEauthorblockA{
% % \textit{dept. name of organization (of Aff.)} \\
% \textit{The Chinese University of Hong Kong}\\
%  Hong Kong SAR, China \\
% jwkang@se.cuhk.edu.hk}
% \and
% \IEEEauthorblockN{2\textsuperscript{nd} Lingwei Meng}
% \IEEEauthorblockA{\textit{The Chinese University of Hong Kong} \\
% Hong Kong SAR, China \\
% lmeng@se.cuhk.edu.hk}
% \and
% \IEEEauthorblockN{3\textsuperscript{rd} Mingyu Cui}
% \IEEEauthorblockA{\textit{The Chinese University of Hong Kong} \\
% Hong Kong SAR, China \\
% mycui@se.cuhk.edu.hk}
% \and
% \IEEEauthorblockN{4\textsuperscript{th} Yuejiao Wang}
% \IEEEauthorblockA{\textit{The Chinese University of Hong Kong} \\
% Hong Kong SAR, China \\
% wangy@se.cuhk.edu.hk}
% \and
% \IEEEauthorblockN{5\textsuperscript{th} Xixin Wu}
% \IEEEauthorblockA{\textit{The Chinese University of Hong Kong} \\
% Hong Kong SAR, China \\
% wuxx@se.cuhk.edu.hk}
% \and
% \IEEEauthorblockN{6\textsuperscript{th} Xunying Liu}
% \IEEEauthorblockA{\textit{The Chinese University of Hong Kong} \\
% Hong Kong SAR, China \\
% xyliu@se.cuhk.edu.hk}
% \and
% \IEEEauthorblockN{7\textsuperscript{th} Helen Meng}
% \IEEEauthorblockA{\textit{The Chinese University of Hong Kong} \\
% Hong Kong SAR, China \\
% hmmeng@se.cuhk.edu.hk}
% }

\maketitle

\begin{abstract}
Multi-talker speech recognition (MTASR) faces unique challenges in disentangling and transcribing overlapping speech.  
To address these challenges,
this paper investigates the role of Connectionist Temporal Classification (CTC) in speaker disentanglement when incorporated with Serialized Output Training (SOT) for MTASR.
Our visualization reveals that CTC guides the encoder to represent different speakers in distinct temporal regions of acoustic embeddings.
Leveraging this insight, we propose a novel Speaker-Aware CTC (SACTC) training objective, based on the Bayes risk CTC framework. 
SACTC is a tailored CTC variant for multi-talker scenarios, it explicitly models speaker disentanglement by constraining the encoder to represent different speakers' tokens at specific time frames.
When integrated with SOT, the SOT-SACTC model consistently outperforms standard SOT-CTC across various degrees of speech overlap. 
Specifically, we observe relative word error rate reductions of 10\% overall and 15\% on low-overlap speech.
% Specifically, we observe relative WER reductions of 10\% overall and 15\% on low-overlap speech.
This work represents an initial exploration of CTC-based enhancements for MTASR tasks, offering a new perspective on speaker disentanglement in multi-talker speech recognition.
The code is available at \textcolor{Blue}{\href{https://github.com/kjw11/Speaker-Aware-CTC}{https://github.com/kjw11/Speaker-Aware-CTC}}.

\end{abstract}

\begin{IEEEkeywords}
multi-talker speech recognition, speech recognition, Connectionist Temporal Classification, cocktail party problem, speech separation
\end{IEEEkeywords}
% \vspace{-5pt}
\section{Introduction}
% \vspace{-2pt}
Natural human conversations always involve multiple speakers, with varying degrees of speech overlaps. 
Multi-talker speech recognition (MTASR) has emerged as a critical field, aiming to transcribe these natural conversational speech. 
While traditional automatic speech recognition (ASR) tasks typically perform monotonic speech-to-text sequence mapping, MTASR presents unique challenges: recognition models are required to disentangle speech from distinct speakers, and separately transcribe their speech.
% In recent years, many approaches have been proposed for speaker-discriminative ASR models.

In recent years, many approaches have been proposed to address this challenge.
% Current mainstream methods can be categorized into two types:
These approaches can be categorized into two types based on their ways of differentiating speakers: branched acoustic encoder (BAE) based models and serialized output training \cite{kanda2020serialized} based models.
BAE models leverage structural priors to disentangle speakers: they separate mixed speech into independent branches, then use shared recognition blocks to transcribe different speakers in parallel.
To align branches with respective speakers, permutation invariant training (PIT)  \cite{yu2017permutation, kolbaek2017multitalker} is applied to calculate ASR loss.
Yu et al. \cite{yu2017recognizing} first adopt a BAE-style model with PIT loss.
Seki et al. \cite{seki2018purely} further extend this approach in a fully end-to-end manner.
Subsequently, Chang et al. \cite{chang2020end} incorporated transformer backbone into this framework.
Further works \cite{lu2021streaming, raj2023surt, tripathi2020end} explored streaming ASR following the BAE framework.
More recently, sidecar separator-based methods \cite{meng2023sidecar, meng23b_interspeech, meng2024empowering} were proposed to convert a single-talker ASR system into a multi-talker one through model-internal separation.
% , improving MTASR by leveraging pre-trained ASR model.

Another line of research lies on Serialized Output Training (SOT) \cite{kanda2020serialized}.
The SOT approach serializes text from different speakers as a single stream, with a special token $\langle sc\rangle$ as a delimiter between speakers.
In contrast to the structural priors of BAE models, this approach relies on attention mechanisms in attention encoder-decoder (AED) \cite{chan2016listen} to disambiguate speakers.
% without explicitly modeling separation.
This confers an advantage in that it does not require pre-defining speakers and branch numbers, allowing it to handle a variable number of speakers.
The superiority of SOT methods has been demonstrated in the M2Met challenges \cite{yu2022m2met, liang2023second}, which provided challenging "in the wild" multi-talker meeting speech.
SOT methods have been further enhanced with speaker information \cite{kanda2020joint, fan2024sa}, time boundary \cite{liang2023ba, kanda2022streaming}, learnable speaker orders \cite{shi2024serialized}, large language models \cite{meng2024large}, and integrated with BAE structures as a hybrid branched SOT model \cite{kang2024cross}.

In contrast to these two paradigms, there is a lack of investigation on the role of connectionist temporal classification (CTC) \cite{graves2006connectionist} in MTASR.
CTC has become a fundamental training criterion for sequence-to-sequence tasks including speech recognition.
Specifically, CTC introduces a blank token to construct alignments between input and target sequences, providing a method to compute posterior probabilities by summing over all possible alignment paths between inputs and target sequences.
Compared to other ASR architecture of AED and Neural Transducer \cite{graves2012sequence, rao2017exploring}, CTC generates all tokens in the sequence simultaneously in a non-autoregressive manner, 
% making generation iterations independent of the target sequence length, 
thus enabling faster decoding speeds.
CTC was also used together with AED models as a joint CTC/Attention model \cite{watanabe2017hybrid}, which has long been considered one of the state-of-the-art approaches for speech recognition.
In the MTASR domain, although the original SOT adopted the AED architecture without including CTC loss, many studies have empirically demonstrated that the joint CTC/Attention SOT model can effectively improve recognition accuracy on overlapped speech \cite{yu2022summary, shen2022volcspeech, ye2022royalflush, liang2023ba}.
However, given CTC's monotonicity assumption, it is counter-intuitive that CTC can non-monotonically map overlapped speech to serialized transcriptions, and there is a lack of literature investigating these results.

In this work, we investigated the effect of CTC, especially when incorporated with SOT.
Our experiments with conformer encoders reveal that CTC loss enables the acoustic encoder to represent different speakers at distinct temporal within the acoustic embeddings.
% The incorporation of CTC significantly improved recognition of mildly overlapped speech.
% When CTC is incorporated with SOT, we observe a significant 20\% relative improvement in recognition accuracy for low-overlap speech.
Distinct from existing BAE and SOT approaches, we attribute CTC's speaker distinction capability to its non-autoregressive reordering capability \cite{chuang2021investigating, shi2023glance}, which is potentially a novel direction for speaker disentanglement in MTASR.
Building on this insight, we proposed a novel speaker-aware CTC (SACTC) as an enhanced and tailored CTC variant for multi-talker scenarios.
This SACTC explicitly models speaker disentanglement by constraining the encoder to represent different speakers' tokens at specific temporal locations.
This is achieved by the Bayes risk CTC framework, where we introduced a speaker-aware risk function to penalty CTC paths with undesired token emit.
In experiments, SACTC was used as an auxiliary loss for SOT-based MTASR models.
Experimental results demonstrate that the SOT-SACTC model consistently outperforms the standard SOT-CTC approach across various degrees of speech overlap.
Notably, we observe word error rate reductions of 10\% overall and of 15\% on low-overlap speech.
To our knowledge, this work represents the first 
exploration of CTC-based enhancements for MTASR tasks.

\section{Methods}
\vspace{-3pt}
\subsection{Revisit CTC in speech recognition}
\vspace{-3pt}
% 1. CTC asr
% 2. CTC理论上怎么完成speaker disentangling?
% 被认为包含monotonic假设，并且某条path domainate over all paths. 这align with ASR, where speech-to-text mapping is monotonic的，输出文本emit aligning with respective pronunciation in speech.
% 3. 
CTC loss guides sequence-to-sequence models by maximizing the posterior probability $p(l|x)$ of the target sequence, where $x=[x_1, ..., x_T]$ represents the input acoustic embedding, e.g., from an acoustic encoder, and $l=[l_1, ..., l_U]$ represents the transcription label sequence.
To compensate for the length discrepancy between $x$ and $l$, CTC introduces blank tokens $\varnothing$ into the label sequence $l$ to construct alignment paths (also as known as CTC labels) $\pi = [\pi_1, ..., \pi_T]$ between $x$ and $l$. $\pi_t$ denotes the output token at time step $t$.
A collapsing function $B(\pi)=l$ maps alignment paths to text labels by collapsing repeated consecutive labels into a single label and removing all blank labels (e.g., $B(\varnothing a \varnothing aabb)=aab$).
Subsequently, the posterior probability $p(l|x)$ of the label sequence can be calculated by summing up the posteriors of all possible alignments:
\begin{equation}
    P(l|x)=\sum_{\pi \in B^{-1}(l)}p(\pi|x)
\label{eq:1}
\end{equation}
where  $\pi \in B^{-1}(l)$ if $B(\pi) = l$. 
$p(\pi|x)$ denotes the posterior probability of path $\pi$, calculated by the product of posterior probabilities of $\pi_t$ cross $T$ time steps:
% \begin{equation}
%     p(\pi|x)=\prod_{t=1}^{T}y_t(\pi_t)
% \end{equation}
% \begin{equation}
% p(\pi|x)=\prod_{t=1}^{T}p(\pi_t|x_t)=\prod_{t=1}^{T}y_t(\pi_t)
% \end{equation}
\begin{equation}
p(\pi|x)=\prod_{t=1}^{T}p(\pi_t|x_t)=\prod_{t=1}^{T}y^t_{\pi_t}
\label{eq:2}
\end{equation}
Here $p(\pi_t|x_t)$ typically consists of linear projection and softmax function to generate frame-wise posterior of $\pi_t$ at the $t$-th frame.
We denote the output as $y^t_{\pi_t}$.

% where $p(\pi_t|x_t)$ generate a frame-wise score of $\pi_t$ at the $t$-th frame, based on softmax function.

% where $p(\pi_t|x_t)$ represents frame-wise score of $\pi_t$ at the $t$-th frame, generated by softmax function.

% where $p(\pi_t|x_t)$ represents frame-wise posterior of $t$-th frame in $x$, generated by softmax function.
% where $y_t$ represents a distribution over all possible tokens for time step $t$, computed by a frame-wise softmax function.
% $y_t(\pi_t)$ stands for the posterior of token $\pi_t$ at the $t$-th frame.

Considering the combinational explosion of permutating all alignment paths, forward-backward algorithm \cite{rabiner1986introduction} is commonly used to calculate $P(l|x)$ effectively. 
First, the original label sequence $l$ is extended by inserting $\varnothing$ symbol between any two non-blank tokens: $l'=[\varnothing,l_1,\varnothing,...,\varnothing,l_U,\varnothing]$.
Then recursively compute the forward-backward variables $\alpha(t, v)$ and $\beta(t, v)$:
\begin{equation}
    \alpha(t,v)=\sum_{\substack{\pi:B(\pi_{1:t})=B(l'_{1:v})\\\pi_t=l'_{v}}} \prod^{t}_{t'=1} y^{t'}_{\pi_{t'}}
\end{equation}
\begin{equation}
    \beta(t,v)=\sum_{\substack{\pi:B(\pi_{t:T})=B(l'_{v:2U+1})\\\pi_t=l'_{v}}} \prod^{T}_{t'=t} y^{t'}_{\pi_{t'}}
\end{equation}
with $(t, v)$ is a node in CTC lattice and $1 \,{\leq}\, t \,{\leq}\, T$, $1 \,{\leq}\, v \,{\leq}\, 2U+1$.
These two variables summarized posterior of all paths passing through node $(t,v)$, with exact $1:v$ prefix and $v:2U+1$ suffix alignment.
Subsequently, for any consent time step $t$, enumerating all possible tokens $v$ in $l'$ will consider all possible paths.
Therefore the CTC posteriors can be calculated by:
\begin{equation}
    P(l|x)=\sum_{\pi \in B^{-1}(l)}p(\pi|x)=\sum^{2U+1}_{v=1}\frac{\alpha(t,v) \cdot \beta(t,v)}{y^t_{l'_v}}
    \label{eq:5}
\end{equation}

\vspace{-4pt}
\subsection{Speaker-aware CTC based on minimizing Bayes risk}
\vspace{-1pt}
\label{sec:sactc}
% Consider a two-talker scenario with serialized target label $l=[l_1^a, ...,l_M^a, l_1^b, ...,l_N^b]$, where $a$ and $b$ stand for 2 speakers.

% Consider a two-talker scenario that the
Consider a two-talker scenario with serialized target label $l=[l_1^a, ...,l_M^a, \langle sc \rangle, l_1^b, ...,l_N^b]$, where $a$ and $b$ stand for 2 speakers and $\langle sc \rangle$ token separates them.
We denote the alignment path as $\pi=[\pi_1^a,..., \pi_K^a, \pi_{K+1}^b,...,\pi_T^b]$, where $\pi_K^a$ represent the last $\langle sc \rangle$ token.
% We note that $K$ is a variable and may have distinct values in different alignment paths.
During CTC training, we consider the case where multi-talker overlapped speech was encoded by an acoustic encoder, resulting in embedding $x$.
Eq. \ref{eq:2} suggests that the information carried by $x$ is inherently encouraged to align with $\pi$.
I.e., $[x_1,...,x_{K-1}]$ encodes speaker $a$ and $[x_{K+1},...,x_T]$ encodes speaker $b$.
We note that the "speaker boundary" $K$ varies across alignment paths, potentially confusing the encoder on how different speakers are represented.
Furthermore, embedding $x$ with a nondeterministic speaker boundary may complicate subsequent processing, e.g., hindering a cascaded ASR decoder from recognizing different speakers.

% We note that the "speaker boundary" $K$ is variable and could have distinct values in different alignment paths.
% Given that CTC loss considers all possible alignments equally, inconsistent $K$ may confuse the front-end acoustic encoder on how different speakers are represented.

Addressing this issue, we propose a speaker-aware CTC training objective as an enhanced and tailored loss function for MTASR.
The core idea is to constrain the encoder model to represent different speakers' tokens at specific time frames, which \textit{explicitly models speaker disentanglement}.  
To control CTC prediction, the Bayes risk CTC (BRCTC) framework \cite{tian2022bayes} was used to introduce preference over alignment paths.
Specifically, BRCTC defined Bayes risk function $r(\pi)$ over posteriors of alignment paths, and the training objective is:
\begin{equation}
    \mathcal{J}_{br}(l,x)=\sum_{\pi \in B^{-1}(l)}r(\pi) \cdot p(\pi|x)
\end{equation}
As paths with the same concerned property could share the same risk value, we group paths by the ending point (frame) of a certain non-blank token and use group-wise risk functions, to control the encoding frames of specific speakers.
% As paths with the same concerned property could share the same risk values, paths can be grouped and use group-wise risk functions. 
% To control the CTC prediction at specific time frames, we group paths with the ending point of a certain non-blank token.
Given a constant non-blank token $l_u=l'_{2u}$, the ending point of $l_u$ is exclusive over time frames $t$, thus CTC posterior can be alternatively calculated by enumerating all possible frames, and the training objective can be reformulated as:
\begin{equation}
    \mathcal{J}_{brctc}(l,x)=\sum^{T}_{t=1}r_g(t)\cdot\frac{\alpha(t,2u) \cdot \hat{\beta}(t,2u)}{y^{t}_{\pi_t}}
    \label{eq:7}
\end{equation}
in which $\hat{\beta}(t,2u)$ is a revised backward variable, summarizing posteriors of the paths where $l_{u}$ ends at $t$, i.e., $t=argmax_\tau$ for $1 \,{\leq}\, \tau \,{\leq}\, T$, s.t. $\pi_\tau=l'_{2u}$.
This can be achieved by eliminating non-ending paths such that $\pi_{t+1} = l'_{2u}$ for $t \,{<}\, T$.
Accordingly:
\begin{equation}
    \hat{\beta}(t, 2u) = 
    \begin{cases}
        \beta(t, 2u) - \beta(t + 1, 2u) \cdot y_{\pi_t}^t & \text{if } t<T \\
        \beta(t, 2u), & \text{Otherwise}
    \end{cases}
\label{eq:8}
\end{equation}
The above grouping strategy inherits the original derivation in \cite{tian2022bayes}.

With Eq. \ref{eq:7} enables computing the training objective by summing over time steps, we define the following speaker-aware risk function to constrain the emitting time of tokens with consideration of their belonging speakers:
% \begin{equation}
%     r_{sa}(s,t)=
%         \begin{cases}
%             \frac{1}{1+e^{-(\lambda(\frac{t}{T}-b))}} & \text{if } s=1 \\
%             \frac{1}{1+e^{(\lambda(\frac{t}{T}+b))}} & \text{Otherwise}
%         \end{cases}, \,\,
%     b=\frac{M}{M+N}
% \end{equation}
\begin{equation}
    r_{sa}(s,t)=
        \begin{cases}
            -\frac{1}{1+e^{(\lambda(\frac{t}{T}-b))}} &
             \text{if } s=1 \\
            -\frac{1}{1+e^{-(\lambda(\frac{t}{T}-b))}} & \text{Otherwise}
        \end{cases}, \,\,
    b=\frac{M}{M+N}
\end{equation}
in which $s$ represents speaker index where speakers were ordered chronologically following the first-in-first-out serialization strategy.
$\lambda$ is an adjustable Bayes risk factor controlling the sharpness of risks, $\lambda=0$ leads to uniform risks for all paths.
And $b$ is a ratio of speaker utterance lengths, used to determine a speaker boundary K invariant to alignment paths.
The $r_{sa}(s,t)$ overall is a conditional Sigmoid function, assigning high or low risks according to the established speaker boundary.
Subsequently, the training objective for a certain $l_u$ is:
\begin{equation}
    \mathcal{J}'_{sa}(l,x,s,u)=\sum^T_{t=1}r_{sa}(s,t) \cdot \frac{\alpha(t,2u) \cdot \hat{\beta}(t,2u)}{y^{t}_{\pi_t}}
\label{eq:10}
\end{equation}
Note that $r_{sa}(s,t)$ is consistently ${\,<}\,0$, so here we minimize expected Bayes risk $\mathcal{J}'_{sa}$, which contrasts with maximizing a posterior as in Eq. \ref{eq:1}.
Furthermore, to constrain every token $l_u \in l$ from all speakers, the final training objective is to minimize the following:
% And the final training objective can be derived by assigning the Bayes risk to all token $l_u \in l$ from all speakers:
% \begin{equation}
%     \mathcal{J}_{sa}(l,x)=\frac{1}{S} \cdot \sum^{S}_{s=1} \,[\, \frac{1}{U_s}  \cdot \sum^{U_s}_{u}log \mathcal{J}'_{sa}(l,x,s,u) \,]
% \end{equation}
\begin{equation}
    \mathcal{J}_{sa}(l,x)=\frac{1}{S} \cdot \sum^{S}_{s=1} \,[\, \frac{1}{U}  \cdot \sum^{U}_{u} \mathbbm{1}(s,u) \cdot log \mathcal{J}'_{sa}(l,x,s,u) \,]
\label{eq:jsa}
\end{equation}
where $S$ is the total speaker numbers and indicator $\mathbbm{1}(s,u)\,{=}\,1$ if token $l_u$ belongs to speaker $s$, otherwise $0$.
Note that if $r_{sa}(s,t)$ provides uniform risks across all paths, $\mathcal{J}_{sa}$ degenerates to vanilla CTC, as it treats all paths equally.
From this perspective, this training objective can be understood as adding a path penalty upon the vanilla CTC objective, where the penalty corresponds to the risk function.

\begin{figure}[t]
\begin{center}
\includegraphics[width=0.9\linewidth,scale=1.00]{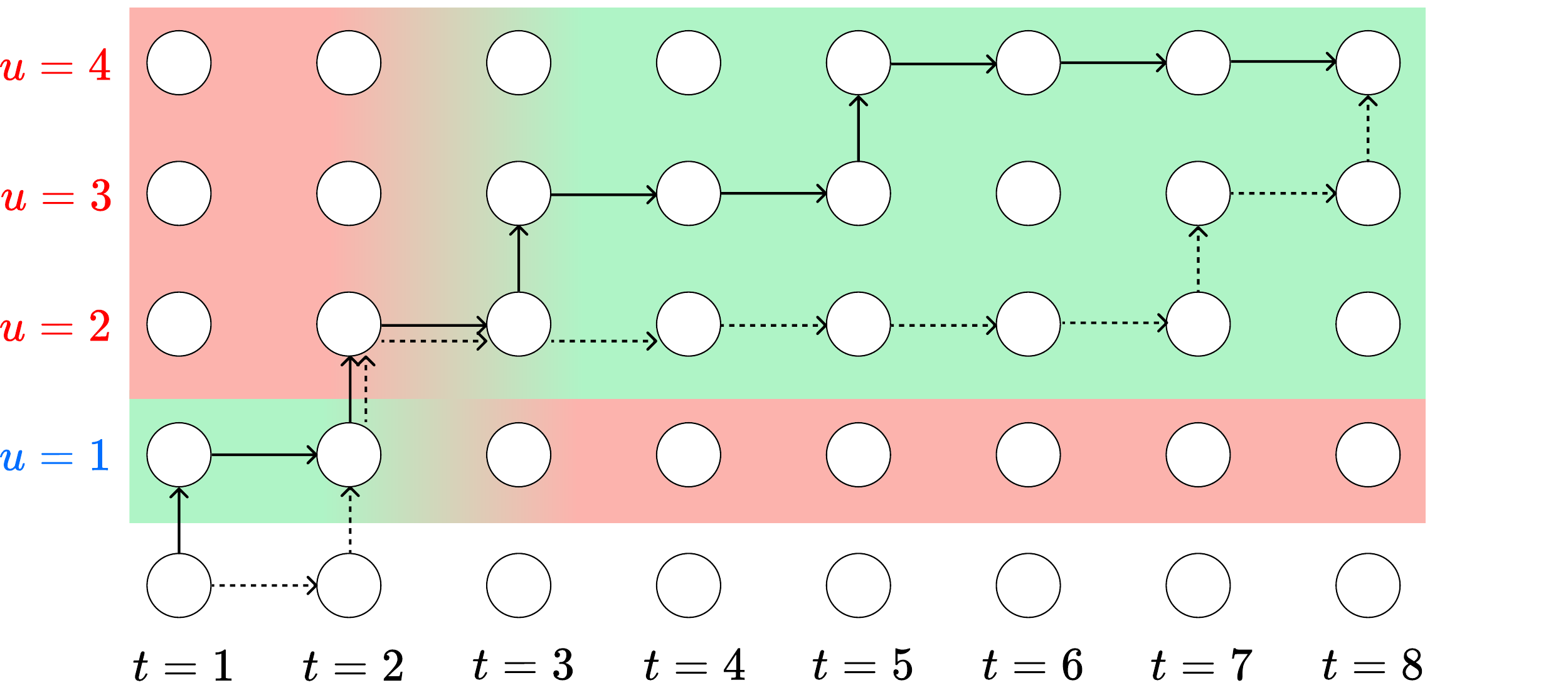}
\end{center}
\vspace{-10pt}
\caption{
A simplified illustration of the proposed speaker-aware risk function with CTC lattice. 
Red area indicates high risk and green for low risk.
Tokens 1 and 2,3,4 are from different speakers.
Two encouraged alignments are shown as examples.
}
\vspace{-10pt}
\label{fig:lattice}
\end{figure}

Fig.~\ref{fig:lattice} presents a simplified illustration of how the proposed training objective works.
It requires the frond-end encoder to disentangle separate speakers onto specific frames.

% This approach was built upon the idea of the Bayes risk CTC (BRCTC) framework.
% Specifically, 

% To achieve this, the Bayes risk CTC (BRCTC) framework } was used to control the CTC prediction, with speaker-dependent risk function

% Addressing this issue, a speaker-aware CTC (SACTC) was proposed to explicitly model speaker disentangling.

% This approach is based on the Bayes risk CTC (BRCTC) framework \cite{tian2022bayes} with tailored speaker-dependent risk functions.
% BRCTC 

% % Addressing this issue, we proposed a speaker-aware CTC to explicitly model speaker disentangling driven by CTC loss.
% This approach is based on the Bayes risk CTC framework \cite{tian2022bayes} with conditionalized 

\section{Experimental setup}
\vspace{-3pt}
% \subsection{Dataset}

\noindent \textbf{Dataset}
Our experiments employed LibriSpeechMix (LSM) \cite{kanda2020serialized} as a benchmark dataset. 
This dataset is derived from the LibriSpeech (LS) \cite{panayotov2015librispeech} corpus, simulated both 2-speaker (LSM-2mix) and 3-speaker (LSM-3mix) mixed speech.
As LSM only provides development and test sets, we generated 2-speaker mixed speeches for training using the similar protocol as in \cite{kanda2020serialized, kang2024cross}.
Specifically, for each sample in the LS 960-hour training set, we randomly sample another sample with a random offset to mix with.
We expect a practical MTASR model can simultaneously handle single- and multi-talker scenarios.
Thus the generated mixed data was combined with the single-talker LS training set, resulting in our training set containing around 560k utterances with 1.7k hours of speech.
To prob model performance on varying degrees of overlapped speech, we further divided the LSM test set into three subsets, representing low, medium, and high overlap conditions.
The corresponding overlap ratios are (0, 0.2], (0.2, 0.5], and (0.5, 1.0] respectively.
The overlap ratio here is defined as the duration of overlaps divided by the total duration of mixed speech.
Besides, we concatenate transcriptions from separate speakers as text labels, using the first-in-first-out serialization strategy.

% \subsection{Model settings}
\noindent \textbf{Model settings}
We implemented CTC and AED ASR models with the ESPnet2 toolkit \cite{watanabe2018espnet}.
For the CTC model, we use a conformer encoder with 12 conformer blocks.
Each block has 4-head self-attention with 256 hidden units and two 1024-dimensional feed-forward layers (macaron style).
The AED model has an additional transformer decoder, comprising 8 transformer blocks with also 4 heads self-attention and 256 hidden units, but a 2048-dimensional feed-forward layer.
As a result, CTC model has 22.14M parameters and AED model has 34.18M parameters.

% \subsection{Training settings and metrics}
\noindent \textbf{Training settings}
The CTC models were trained with vanilla CTC or proposed speaker-aware CTC (SACTC) objectives.
And AED model was trained with sole AED loss (w/o CTC) or joint-CTC/attention loss, where CTC weight was set as 0.3.
During training, Adam optimizer was used with learning rate of 5e-4, warm-up steps of 25k, and batch bins of 35M.
Our preliminary study shows CTC converges slower than AED in MTASR, thus we trained CTC model for 80 epochs, while AED models for 50 epochs.
After training, the best 10 epochs on the dev set were fused as the final models.

\noindent \textbf{Metrics}
For single-talker condition, we used the standard word error rate (WER) as the evaluation metric.
For multi-talker condition, we deployed permutation-invariant WER, a common metric for SOT approaches.
This metric compares all possible permutations of speaker orders and picks up the lowest WER.
Additionally, we also implemented overlap-aware WER (OA-WER) \cite{kang2024cross}.
OA-WER averages WERs across various overlap ratios, aiming to balance the impact of different degrees of overlapped speech.

% \vspace{-5pt}
% \input{tables/main_res3}

\setlength{\tabcolsep}{3pt}
\begin{table}[t]
\vspace{-7pt}
\centering

\caption{
WER (\%) of vanilla CTC and SACTC in MTASR. C1 and E1 are the main experiments.
"dec." stands for decoding.
}
\vspace{-8pt}
\label{tab:t1}
\scalebox{0.85}{
\renewcommand{\arraystretch}{1.3} % line hight
\begin{tabular}{l|l|cc|cc|cccc}
\bottomrule

\multirow{3}{*}{\textbf{ID}} & \makecell[c]{\multirow{3}{*}{
\textbf{System}}}  &\multicolumn{2}{c|}{\textbf{Librispeech}} &\multicolumn{6}{c}{\textbf{LibrispeechMix-2mix}} \\
 \cline{3-4} \cline{5-10} 
& & \multirow{2}{*}{\textbf{dev}} & \multirow{2}{*}{\textbf{test}} & \multirow{2}{*}{\textbf{Dev}} & \multirow{2}{*}{\shortstack{\textbf{Test}\\(Overall)}}& \multicolumn{4}{c}{\textbf{Test} (Conditional)} \\
\cline{7-10}
% & & & & & & \textbf{(0, 20]} & \textbf{(20, 50]} & \textbf{(50, 100]} & \textbf{Average} \\
& & & & & & low & mid & high & OA-WER \\
% & & & & & & low & median & high & OA-WER \\
\noalign{\hrule height 0.9pt}
A1 & SOT & 4.1 & 4.6 & 7.9 & 9.2 & 9.0 & \hl{8.0} & 12.8 & 9.9 \\
\hline
B1 & CTC & 5.0 & 5.4 & 11.7 & 11.1 & 7.5 & 12.4 & 18.2 & 12.7\\
\hline
\hl{C1} & SOT+CTC & 4.3 & 4.5 & 8.4  & 8.8 & 7.1 & 9.0  & 13.1 & 9.7\\
C2 & \, $\hookrightarrow$ AED only dec. & 4.8 & 5.4 & 11.5 & 12.9 & 11.7 & 12.5 & 17.3& 13.8\\
C3 & \, $\hookrightarrow$ CTC only dec. & 5.5 & 5.6 & 12.7 & 12.0 & 7.8 & 13.8 & 19.7 & 13.8\\
\hline
D1 & SACTC & 5.4 & 5.6 & 13.5 & 12.3 & 8.0 & 13.9 & 20.8 & 14.2\\
\hline
\hl{E1} & SOT+SACTC & 3.9 & \hl{4.1} & 8.2 & \hl{8.0} & \hl{6.0} & 8.4 & 12.8 & \hl{9.1} \\
E2 & \, $\hookrightarrow$ AED only dec. & 4.0 & 4.5 & 8.4 & 8.8 & 8.2 & \hl{8.0} & \hl{12.3} & 9.5\\
E3 & \, $\hookrightarrow$ CTC only dec. & 5.5 & 5.8 & 12.5 & 11.9 & 8.1 & 13.1 & 19.9 & 13.7\\

\toprule
\end{tabular}
}
\vspace{-10pt}
\end{table}

\section{Results and discussions}
\vspace{-3pt}
In this section, we first analyze the effect of vanilla CTC in SOT-based MTASR.
We then present and discuss the experimental results of the proposed SACTC approach, comparing it to vanilla CTC.

\vspace{-3pt}
\subsection{Analysis of vanilla CTC}
\vspace{-3pt}
Previous research has demonstrated that integrating CTC with SOT improves MTASR performance.
% Integrating CTC with SOT has been shown to improve MTASR performance.
% Many studies demonstrated that incorporating CTC with SOT improves MTASR. 
We reproduced these experiments, with results presented in Table~\ref{tab:t1}, systems A to C.
Comparing CTC with SOT, we observe that while CTC generally performed worse, it achieved better WER on the low-overlap subset (7.5 vs. 9.0).
Moreover, incorporating CTC with SOT (C1) didn't enhance single-talker performance but improved multi-talker recognition, particularly on the low-overlap subset (9.0→7.5).
% Second, SOT+CTC model overall improves sole SOT model on LSM-2mix (9.2→8.8), particularly on the low-overlap subset (9.0→7.5).
However, \textit{the addition of CTC led to decreased performance on mid- and high-overlap speech.}
These results validate that CTC could assist in recognizing low-overlap speech, while it degrades performance when encountering more severe overlaps.
% Besides, decoding with AED only (C2) performs quite poorly on low-overlap speech, but fixed significantly when jointly decoding with CTC (C1).

\begin{figure}[h]
\begin{center}
\includegraphics[width=1\linewidth,scale=1.00]{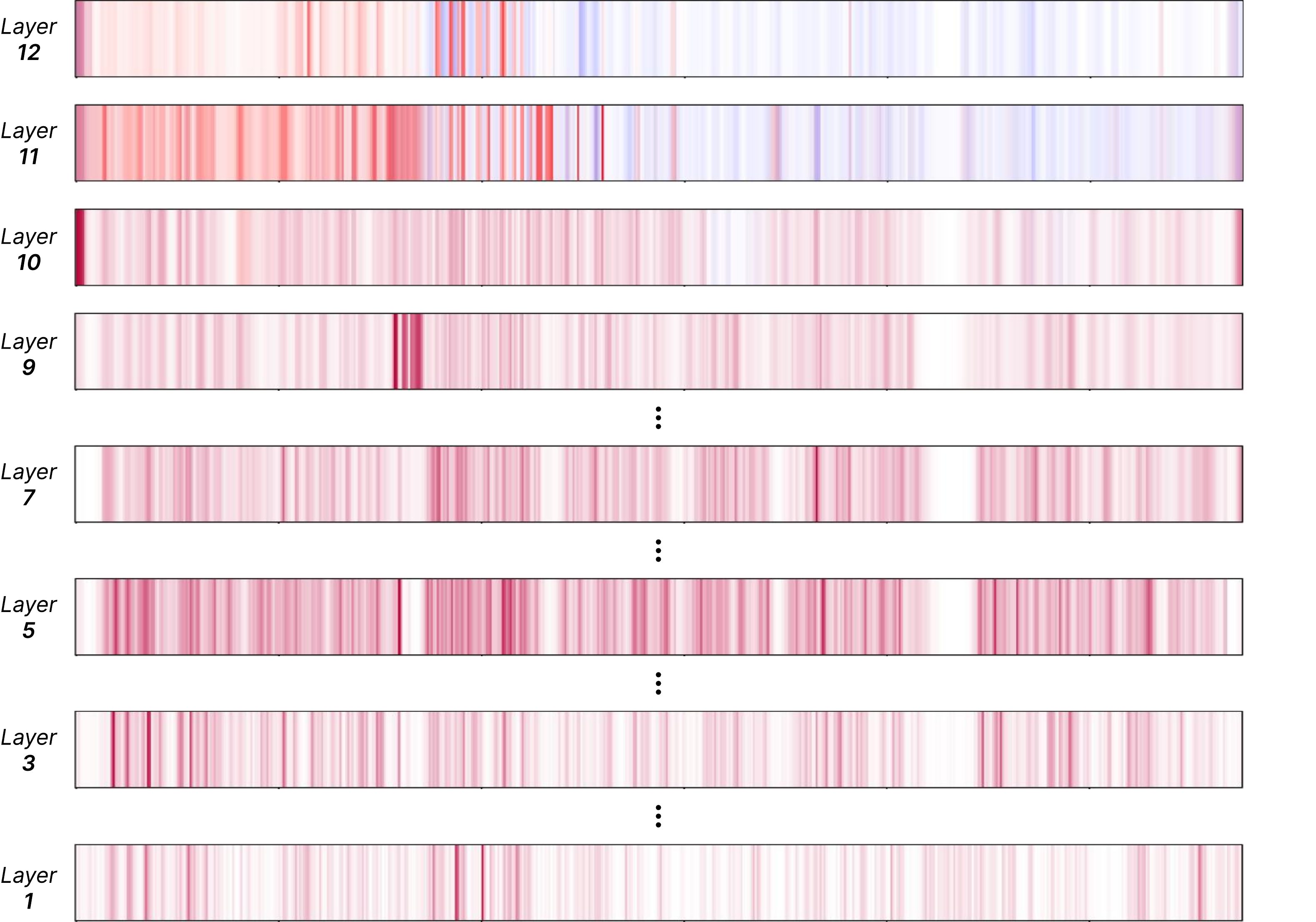}
\end{center}
\vspace{-10pt}
\caption{
Visualization of top-50 attended frames for two speakers (red and blue colors).
Purple colors represent two speakers attending simultaneously.
}
\vspace{-0pt}
\label{fig:layers}
\end{figure}

\begin{figure}[h]
\begin{center}
\vspace{-5pt}
\includegraphics[width=0.9\linewidth,scale=1.00]{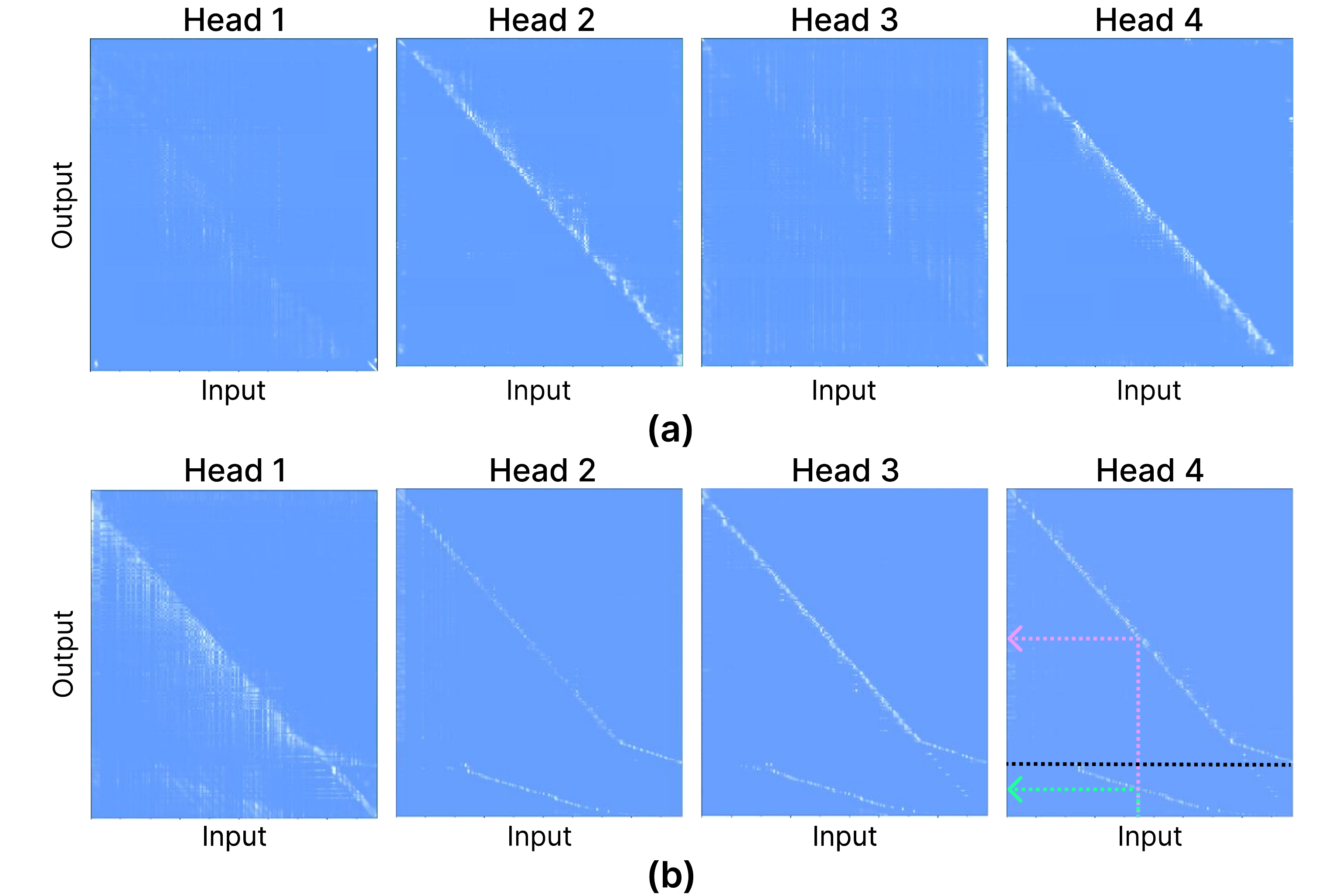}
\end{center}
\vspace{-15pt}
\caption{
Attention matrices in the last conformer blocks of SOT (a) and SOT-CTC (b) models.
In (b), the overlapped area was encoded into separate output frames.
}
\vspace{-15pt}
\label{fig:attn}
\end{figure}

To better understand the interaction of CTC and multi-talker speech, we examined the attention patterns in the conformer encoder for different speakers.
In detail, we visualized the top 50 attended frames in self-attention for each CTC-emitted token, then accumulated attentions for tokens from two speakers in distinct colors.
Fig.~\ref{fig:layers} illustrates an interesting pattern: two speakers generally attended all frames in shallower blocks, while from layer 10 onwards, two speakers began to focus on distinct regions.
This aligns with our derivation in Section~\ref{sec:sactc}.
Notably,  we did not observe this phenomenon in the sole SOT system.
Fig.~\ref{fig:attn} further visualize the attention matrices in the last conformer block.
Compared to the sole SOT system, the use of CTC leads to information re-ordering: certain portions of the input embedding are attended to distinct regions of the output embedding (illustrated in Fig.~\ref{fig:attn}(b) Head 4). 
Moreover, these repeatedly attended regions show a direct correlation to the overlapped area in the input speech.
These visualizations suggest that with CTC guidance, the self-attention modules disentangle different speakers along the time dimension to align with concatenated labels.
Together with WER results, we hypothesize that this capacity is limited for handling severely overlapped speech.

\vspace{-3pt}
\subsection{Performance of SACTC}
\vspace{-3pt}
We evaluated the proposed SACTC approach using a default risk factor parameter of 15. 
Initially, we trained a model with the SACTC objective alone. 
As shown in Table~\ref{tab:t1}, SACTC by itself did not outperform the vanilla CTC model (B1 vs. D1). 
However, when combined with SOT (D1), the model showed significant improvements over vanilla CTC: overall LSM-2mix WER improved from 8.8 to 8.0, and mid-overlap WER from 12.4 to 8.4.
This outcome is understandable, as SACTC is designed to enhance MTASR embedding with deterministic speaker disentangling, thus not necessarily improving token-level recognition\footnote{
For encoder-only models, there might exist potential trade-offs between these two aspects.
}.
When integrated with SOT, SACTC enhanced low-overlap recognition similar to vanilla CTC, while mitigating performance degradation on more severe overlaps.
Experiment E2 also supports this interpretation.
For the model trained with SOT+SACTC, AED-only decoding led to performance gains compared to SOT+CTC, particularly improving recognition in high-overlap conditions (13.1→12.3). 
This suggests that SACTC resulted in embeddings with enhanced speaker discriminability.
Based on these findings, we propose that AED-only decoding should be preferred for heavily overlapped scenarios.

\begin{table}[t]
\centering

\caption{
WER (\%) of SOT+SACTC with different risk factors, where 15 is the default setting.
}
\vspace{-8pt}
\label{tab:t3}
\scalebox{0.85}{
\renewcommand{\arraystretch}{1.3} % line hight
\begin{tabular}{l|c|cc|cc|cccc}
\bottomrule

\multirow{3}{*}{\textbf{ID}} & \multirow{3}{*}{\textbf{Risk factor}}  &\multicolumn{2}{c|}{\textbf{Librispeech}} &\multicolumn{6}{c}{\textbf{LibrispeechMix-2mix}} \\
 \cline{3-4} \cline{5-10} 
& & \multirow{2}{*}{\textbf{dev}} & \multirow{2}{*}{\textbf{test}} & \multirow{2}{*}{\textbf{Dev}} & \multirow{2}{*}{\shortstack{\textbf{Test}\\(Overall)}}& \multicolumn{4}{c}{\textbf{Test} (Conditional)} \\
\cline{7-10}
% & & & & & & \textbf{(0, 20]} & \textbf{(20, 50]} & \textbf{(50, 100]} & \textbf{Average} \\
& & & & & & low & mid & high & OA-WER \\
% & & & & & & low & median & high & OA-WER \\
\noalign{\hrule height 0.9pt}
% D1 & SOT+BRCTC & 3.9 & 4.1 & 8.2 & 8.0 & 6.0 & 8.4 & 12.8 & 9.1 \\
% \hline
C1 & SOT+CTC & 4.3 & 4.5 & 8.4  & 8.8 & 7.1 & 9.0  & 13.1 & 9.7\\
\hline
F1 & 5 & 4.3 & 4.6 & 8.4 & 8.8 & 7.3 & 8.9 & 12.7 & 9.6\\
F2 & 10 & 4.0 & 4.4 & \hl{8.1} & 8.3 & 6.5 & 8.6 & \hl{12.4} & 9.2 \\
F3 & 15 & \hl{3.9} & \hl{4.1} & 8.2 & \hl{8.0} & \hl{6.0} & \hl{8.4} & 12.8 & \hl{9.1} \\
F4 & 20 & 4.1 & 4.3 & \hl{8.1} & 8.3 & 6.4 & 8.6 & 12.7 & 9.2\\

\toprule
\end{tabular}
}
\vspace{-12pt}
\end{table}

\noindent \textbf{Impact of hyperparameter}
We also investigated the impact of different risk factors (RFs) in SOT-SACTC, as shown in Table~\ref{tab:t3}.
It shows that all tested RFs yielded improvements over the baseline C1. 
With a small RF of 5, the performance was close to the baseline, while the best result was achieved using RF=15.

\noindent \textbf{Generalize to more speakers}
A key advantage of SOT-based models is their ability to generalize to a greater number of speakers than present in the training data.
To test this, we evaluated our models on the LSM-3mix test set, despite all models being trained on 1 and 2 speaker scenarios.
As shown in Table~\ref{tab:3spkr}, the trends observed were similar to those in the 2-speaker scenario. 
The SOT-CTC model significantly outperformed the SOT model on low-overlap speech (17.7\% vs. 23.6\%), but showed degraded performance on high-overlap speech (30.1\% vs. 29.5\%). 
In contrast, the SOT-SACTC model achieved the best performance across all conditions.

% These experiments demonstrate the effectiveness of the proposed SACTC as a multi-talker-enhanced loss function, which significantly improved SOT-based MTASR models.

% \vspace{-10pt}
\begin{table}[htbp]
\centering
\vspace{-7pt}
\caption{
WER (\%) of MTASR models in 3-speaker test set.
}
\vspace{-8pt}
\label{tab:3spkr}
\scalebox{0.85}{
\renewcommand{\arraystretch}{1.3} % line hight
\setlength{\tabcolsep}{4pt}
\begin{tabular}{l|cc|cccc}
\bottomrule

\multirow{2}{*}{\textbf{System}} & \multirow{2}{*}{\textbf{Dev}} & \multirow{2}{*}{\shortstack{\textbf{Test} \\ (Overall)}}& \multicolumn{4}{c}{\textbf{Test} (Conditional)} \\
% sys& dev& a & b& c & d & e \\
\cline{4-7}
 &  &  & low & mid & high & OA-WER \\
\noalign{\hrule height 0.9pt}
SOT & 22.9 & 25.3 & 23.6 & 24.3 & 29.5 & 25.8 \\
SOT-CTC & 23.5 & 23.6 & 17.7 & 23.3 & 30.1 & 23.7 \\
SOT-SACTC & \hl{22.6} & \hl{22.6} & \hl{15.9} & \hl{22.7} & \hl{29.1} & \hl{22.6} \\

\toprule
\end{tabular}
}
\end{table}
\vspace{-10pt}
% \vspace{-5pt}
\section{conclusions}
\vspace{-3pt}

In this work, we investigated the effect of CTC in multi-talker speech recognition (MTASR) based on serialized output training (SOT). 
Our findings reveal that the CTC training objective guides the ASR encoder to encode different speakers into distinct temporal regions within acoustic embeddings.
Building upon this insight, we leveraged the Bayes risk CTC framework and proposed a speaker-aware CTC (SACTC), an enhanced CTC variant tailored for MTASR.
The core idea of SACTC is to constrain the encoder model to represent different speakers' tokens at specific time frames, explicitly modeling speaker disentanglement.
SACTC was used as an auxiliary loss for SOT-based MTASR models in our experiments.
Experimental results demonstrate that the SOT-SACTC model consistently outperforms the standard SOT-CTC approach across various degrees of speech overlap.
Notably, we observe relative WER reductions of 10\% overall and of 15\% on low-overlap speech.
To our knowledge, this work represents the first 
exploration of CTC-based enhancements for MTASR tasks. 
Future research directions may include extending SACTC to streaming seniors and exploring its potential in non-autoregressive speech recognition. 

\section{Acknowledgements}
\vspace{-3pt}
This work is supported by the HKSARG Research Grants Council’s Theme-based Research Grant Scheme (Project No. T45- 407/19N) and the CUHK Stanley Ho Big Data Decision Research Centre.

\newpage
\bibliographystyle{IEEEtran}
\bibliography{IEEEfull,myRef}

\end{document}